\documentstyle[12pt]{article}

\newcommand{\ii }{\'{\i}}

\newtheorem{theorem}{Theorem}

\newcommand{\rf}[1]{(\ref{eq:#1})}
\newcommand{\be}{\begin{equation}}
\newcommand{\te}{\end{equation}}
\newcommand{\C}{{C\!\!\!\!1}}
\newcommand{\R}{I\!\! R}

\newcommand{\Y}{{\cal Y}}
\newcommand{\N}{{\cal N}}

\newcommand{\M}{{\cal M}}

\newcommand{\ud}{\underline}

\begin{document}

\author{ M. D. Maia\thanks{E-Mail: maia@mat.unb.br}\\
Universidade de Bras\'{\i}lia, Departamento de Matem\'{a}tica\\
Bras\ii lia, DF. 70910-900\\
and\\
E. M.  Monte\\
Universidade de Bras\'{\i}lia, Departamento de Matem\'{a}tica\\
Bras\ii lia, DF. 70910-900 and\\
Universidade Federal  da Para\ii ba, Departamento de Matem\'atica\\
Campina Grande, Pb. 58109-000}

\title{\hfill \mbox{{\small UnB.FM -003/94}} \\
\vspace{2cm} The Signature Problem for Embedded Space-times }

\maketitle

\begin{abstract}
The compatibility between general relativity and  the  property
that  space-times are  ebeddable manifolds is  further examined.
It is shown that the signature of the embedding space is
uniquely determined provided the embedding space is real and its dimension is
kept to the minimal.  Signature changes produce  complex embeddings which in
turn may induce topological changes in the space-time.
Space-time  signature preserving symmetries identify the twisting
vector as a real connection  on space-time whose curvature is described by the
Ricci's equation in terms of the second fundamental form.
\end{abstract}

\newpage

\section{Introduction}

The four-dimensionality of space-time is deeply rooted in experimental
facts. On such grounds there is no direct evidence to support the idea of a
higher dimensional physical space, at least with today's available level of
high energy physics ($\approx 10^{3} $ GeV). Still, the hypothesis of higher
dimensional of space-time ${\cal M}_{D}$ appears to be consistent with the
extra degrees of freedom required by some unified theories. The two best known
examples of geometric unification  are based on higher dimensions.
Kaluza-Klein theory, mimics general relativity to the extent that
it uses the same Einstein-Hilbert action with the metric as the dynamical
variable. On the other hand, string theory appeals to the notion of a
minimal manifold, using the Nambu-Goto action with the embedding coordinates
as the dynamical variables.

In those high dimensional models, supposedly all dimensions were once, all
accessible. However, at some later time, $D-4$ of
those dimensions become invisible so to speak, to any observer with ''low
energy'' probes (anything below Planck's energy: $10^{19}$GeV). Such
''dimensional reduction'' is usually explained by the ingenious spontaneous
compactification mechanisms ${\cal M}_D\rightarrow V_4\times B_n$ where $V_4$
is the four dimensional space-time and $B_n$ is some compact internal space
with distinct characteristics in each theory \cite{Niew:1}.
In these particular constructions, there are no differential-geometric
constraints imposed on the way  $V_{4}$ and  $B_{n}$  are put together,
resulting that the compact internal spaces play little role in the classical
dynamics of $V_{4}$.

Another, perhaps more natural, way of introducing  higher dimensions is to
use the  fact that  all manifolds, including space-times are   embeddable into
some higher dimensional space  $\M_{D}$  \cite{Joseph:1}, \cite{Ne'emman:1}.
However, it is not clear that this mathematical property is compatible
with the  physics of the space-times. There  are at least two  basic problems
which must be solved before we can make  good use of the  space-time
embeddings.  The first problem (the signature problem) referes to the
existence of  different embedding signatures  for the same space-time.
The other problem refers to the physical  role, if any,  of the extrinsic
curvature.

The  number of dimensions $D$  depends on the differentiable nature of the
embedding functions. If we agree with Janet and Cartan that those functions are
analytic, then the embedding space has dimension  $D\le d(d+1)/2
$ \cite{JC:1}. However, analytic functions may be too special as
compared to differentiable functions to describe high energy physics. If we
assume the more likely differentiable embedding, the  number of
dimensions becomes $D \le d(d+3)/2$ \cite{Greene:1}. Of course, in most known
situations we need less than those limits and it makes sense to
adopt a principle of economy of dimensions: If a given space-time has been
proven to be embeddable in D dimensions, then we will not use more than D
dimensions.
Furthermore,  the fact that it is always possible to find
an embedding in a flat space, we will assume that our embedding space is flat.
However, we  admit that  physics may not act that way. As in the Kaluza-Klein
model,  it is possible that the embedding space is dynamical with a curvature
and topology that changes with time.

Consider a space-time $V_{4}$ with metric $g_{ij},$ solution of Einstein's
equations and its local isometric embedding in a flat D-dimensional manifold
${\cal M}_{D}$. That is, a 1:1 map
$$
{\cal Y} :V_{4}\rightarrow {\cal M}_{D}
$$
such that  \cite{Fri:1}
\begin{equation}
\label{eq:IMB}g_{ij} ={\cal Y}^{\mu}{}_{,i}{\cal Y}^{\nu}{}_{,j}{\cal G}
_{\mu\nu},\;\;{\cal N}^{\mu}_{A}{\cal Y}^{\nu}{}_{,i}{\cal G}
_{\mu\nu}=0,\;\; {\cal N}^{\mu}_{A}{\cal N}^{\nu}_{B}{\cal G}
_{\mu\nu}=g_{AB}=\epsilon_{A}\delta_{AB}
\end{equation}
where $x^{i}$ are coordinates in space-time and ${\cal N}_{A}$ are  $D-4$
vector fields orthogonal to the embedded space-time
\footnote{Lower case Latin indices run from 1 to 4 and capital Latin
indices run from 5 to D, where D is the smallest possible embedding dimension.
All Greek indices run from 1 to D. The indicated antissymetrization applies
only to the indice of the same kind near the brackets.}
Here $\epsilon_{A}=\pm 1$ and ${\cal G}_{\mu\nu}$ denote the components of
the metric of ${\cal M}_{D}$ in the embedding coordinates ${\cal Y}^{\mu}$.
If we prefer, we may use Cartesian coordinates where the metric of $\M_{D}$
has components $\eta_{\mu\nu}$.

The embedding functions $\Y^{\mu}(x^{i}) $ can be obtained by
integrating the Gauss and Weingarten equations:
\begin{eqnarray}
{\cal Y}^{\mu}_{;ij} & =& g^{MN}b_{ijM}{\cal N}^{\mu}{}_{N} \label{eq:Gauss} \\
{\cal N}^{\mu}_{A\; ;j} & = & - g^{mn}b_{jmA}{\cal Y}^{\mu}{}_{,n}
+g^{MN}A_{jAM}{\cal N}^{\mu}{}_{N}  \label{eq:Weingarten}
\end{eqnarray}
where $b_{ijA}$ are the components of the second fundamental form and $
A_{iAB}$ are the components of the twisting vector. Since ${\cal M}_D$ is
flat we may always choose ${\cal Y}^\mu $ as Cartesian coordinates we obtain
explicitly
\begin{equation}
\label{eq:b}b_{ijA}=-{\cal Y}^\mu {}_{,i}{\cal N}^\nu {}_{A,j}{\eta}_{\mu
\nu }={\cal Y}^\mu {}_{,ij}{\cal N}_A^\nu {\eta}_{\mu \nu }
\end{equation}
It follows that $b_{ijA}$ is symmetric in the first two indices. Likewise,
the expression of the twisting vector is:
\begin{equation}
\label{eq:a}A_{iAB}={\cal N}^{\mu}{}_{A} {\cal N}^{\nu}{}_{B,i} {\eta}_{\mu
\nu }
\end{equation}
so that $A_{iAB}=-A_{iBA}$. These two quantities, determine completely the
extrinsic geometry of the space-time, giving a  measure of the local shape of
the space-time as compared to the tangent space. Obviously, if the
embedding of the space-time is given by the embedding coordinates as for
example in \cite{Rosen:1}, then all we have to do is to calculate
$\N^{\mu}{}_{A}$, $b_{ijA}$ and $A_{iAB}$ \cite{Roque:1}. However, if we do so
we learn very little over  what we already know from the intrinsic geometry.
The situation may be different if we assume that the  embedding is not known
but that it results from the space-time dynamics.
Since in general $b_{ijA}$ and $A_{iAB}$ are independent of the
metric, we may express this dynamics in terms of those variables instead of
the embedding coordinates ${\cal Y}^\mu $. The integrability conditions for
(\ref{eq:Gauss}) and (\ref{eq:Weingarten}) are the well known
Gauss-Codazzi-Ricci equations for sub manifolds which may be written as
\begin{equation}
\label{eq:GCR}
\begin{array}{lll}
R_{ijkl}=2g^{MN}b_{i[kM}b_{j]lN},\vspace{2mm} \\ \vspace{2mm}
b_{i[jA;k]}=g^{MN}A_{[kAM}b_{j]iN}, \\ \vspace{2mm}
A_{[jAB;k]}+g^{MN}A_{[jMA}A_{k]NB}=-g^{mn}b_{m[jA}b_{k]nB}
\end{array}
\end{equation}
There are some specific procedures for integrating these equations, as for
example in \cite{Collinson:1},\cite{Matsumoto:1}. In the next section  we look
at the signature problem and  in section 3  we  deal with the interpretation of
the  twisting vector as a gauge field.

\section{The Signature Problem}
Assuming that the space-time has Lorentz signature, then the embedding space
has necessarily a pseudo Euclidean signature, possibly with several
time-like directions, one of them necessarily lying on the tangent plane.
It is possible to find different embedding signatures for the
same space-time. If the extrinsic properties of the space-time are to be
physically relevant, then such ambiguity is not acceptable.
The mentioned principle of economy of dimensions has the following
consequence:

\begin{theorem}
IF $D$ is the smallest dimension in which we can isometrically embed a
non-flat space-time $V_4$ in a real space ${\cal M}_D$, then the signature
of this space is unique.
\end{theorem}

Suppose that we have two embeddings of the same space-time ${\cal Y}:
V_{4}\rightarrow {\cal M}_{D}$ and ${\cal Y}^{\prime}:{\cal V}_{4}
\rightarrow {\cal M}^{\prime}_{D}$ which differ only in signature: $p-q$ in
the first case and $p^{\prime}-q^{\prime}$ in the second case. Since the
tangent spaces to $V_{4}$ have the same Minkowski signature, without loss of
generality they may be identified. That is, we may define a
map $ T: {\cal M}_{D}\rightarrow {\cal M^{\prime}}_{D}$ such that its
derivative  $T^{*}$ restricted to the tangent space $TV_{4}$ is the identity:
$T^{*}\rfloor_{TV_{4}}=T^{*}_{1}=Id$. On the other hand, the restriction
$T^{*}_{2}$ to the subspaces $V_{4}^{\perp}$ of  $\M_{D}$ orthogonal to the
space-time, $T^{*}\rfloor_{TV_{4}^{\perp}}$ is a general linear transformation
(figure
1). In terms of the embedding coordinates and normal vectors this is
equivalent to
\begin{equation}
{\cal Y'}^{\mu}{}_{, i}= {\cal Y}^{\mu}{}_{, i},\;\;\;\;\mbox{and}
\;\;\;\; {\cal N}^{\prime}_{B} =T^{A}{}_{B} {\cal N}_{A}
\end{equation}
\begin{figure}[tbh]
\begin{picture}(50,50)(0,0)
\thinlines
\put(45,31){\vector(2,1){25}}
\put (45,28.){\vector(2,-1){25}}
\put(77,43){\vector(0,-1){26}}
\put(94,43){\vector(0,-1){26}}
\put(107,43){\vector(0,-1){26}}
\it
\put(39,29){$V_{4}$}
\put(75,13){$\M'_{D}\longrightarrow T{V'}_{4}\oplus  T{V'}_{4}^{\perp}$}
\put(75,45){$\M_{D}\longrightarrow {TV}_{4}\oplus{TV}_{4}^{\perp}$}
\put(55,39){$\Y$}
\put(55,18){${\Y'}$}
\put(79,29){$T$}
\put(99,29){$T^{*}_{1}$}
\put(111,29){$T^{*}_{2}$}
\put(84,50){$\pi$}
\put(84,9){$\pi '$}
\label{fig:T}
\put(55,0){\small Fig.1: Two embeddings of $V_{4}$}
\end{picture}
\end{figure}
{}From (\ref{eq:b}), the second fundamental form transforms as
$b^{\prime}_{ijB}= T^{A}{}_{B}b_{ijA}  $ so that
for the second embedding Gauss equation  is
\begin{equation}
\label{eq:G'}R_{ijkl} = 2g'^{MN}b^{\prime}_{i[kM}b^{\prime}_{l]jN},
\end{equation}
Comparing with Gauss' equation in (\ref{eq:GCR}) we obtain
\begin{equation}
\label{eq:GG}2\sum_{A=5}^{D}{g''}^{AB}b_{i[kA}b_{l]jB}=0,
\end{equation}
where we have denoted
\begin{equation}
{g''}^{AB}=g^{AB}-g'^{MN}T^{A}{}_{M}T^{B}{}_{N}. \label{eq:ggg}
\end{equation}
It remains to see if (\ref{eq:GG}) admits a non trivial solution ${
g''}^{AB}$ of the form $\epsilon''_{A}\delta_{AB}$
, with $\epsilon''^{A} =\pm 1$. We have the following
possibilities\\

a) All ${g''}^{AB}$ coincide with $g^{AB}$: \vspace{1mm}\\ ${
g''}^{AB}=g^{AB}\;\;\forall\;\;A,B=5,...,D. $ In this case we have
$g'^{MN}T^{A}_{M}T^{B}_{N}=0$ which is not possible since the left
hand side of equation (\ref{eq:G'}) becomes identically zero, contradicting
the hypothesis of a non flat $V_{4}$.

b) Only some values of ${g''}^{AB}$ coincide with $g^{AB}$.\\
For example, suppose that ${g''}^{AB} \neq g^{AB}$ for $A,B
=5,...,D_{1}$ and ${g''}^{AB} = g^{AB}$ for $A,B =D_{1}+1, ...,D,$
where $5< D_{1}< D $. From  (\ref{eq:GG}), it follows that
$$
2\sum _{A,B=5}^{D_{1}} g''^{AB}b_{i[kA}b_{l]jB} +
2\!\!\sum_{A,B=D_{1}+1}^{D}g^{AB}b_{i[kA}b_{l]jB}=0.
$$
Therefore, replacing the last term in the Gauss equation of (\ref{eq:GCR}),
we get
$$
R_{ijkl}=2\sum_{A,B=5}^{D_{1}} g^{AB}b_{i[kA}b_{l]jB}
+2\sum_{A,B=D_{1}+1}^{D} g^{AB}b_{i[kA}b_{l]jB} =
2\sum_{A,B=5}^{D_{1}}(g^{AB}-g''^{AB})b_{i[kA}b_{l]jB}.
$$
Since, the quadratic form $g^{AB}-g''^{AB}$ can always be
diagonalized, we may write
$g^{AB}-g''^{AB}=g'''^{AB}=\epsilon^{\prime\prime\prime}_{A}\delta^{AB}$.
Therefore the last equation corresponds to Gauss' equation for a third
embedding of $ V_{4}$ in a space with $D_{1}$ dimensions, contradicting
the hypothesis.

c) The  remaining possibility corresponds to a trivial solution
${g''}^{AB}=0,\; \forall \; A,B=5,..,D$. In other words,
\begin{equation}
\label{eq:T}g^{AB} = T^{A}{}_{M} g'^{MN}T_{N}{}^{B} .
\end{equation}
In matrix notation, $g=Tg^{\prime}T^{t}$, so that $(\mbox{det} T)^{2} =
\mbox{det} g/\mbox{det} g^{\prime}$. Therefore, we have $\mbox{det}(T) = \pm
1, \,\mbox{or} \, \mbox{det}(T) = \pm i$.

In the cases $det (T)= \pm 1$ the
signature of the embedding space remains unchanged. In particular when $det
(T)=1$, $T$ belongs to the group of pseudo rotations of the normal vectors $
{\cal N}^{\mu}{}_{A}$. Since the tangent space to $V_{4}$ has  Minkowski
signature and
$\M_{D}$ has signature $p-q$, this group is $SO(p-3,q-1)$.

On the other hand if $\mbox{det}(T)=\pm i$ we have different signatures
corresponding to a complex $T$. A complexification ${\cal M}_D/{\C}$  of
$\M_{D}$ is defined by a pair of maps \cite{Tra:1}:
$$
\left\{
\begin{array}{l}
+:{\cal M}_D\times {\cal M}_D\longrightarrow {\cal M}_D\times {\cal M}_D\;\;
\mbox{given by}\;\;(u,v)+(w,x)=(u+w,v+x)\vspace{3mm}\\
\; {*} :{\cal M}_D\times {\cal M}_D\longrightarrow {\cal M}_D\times
{\cal M}_D\;\;\mbox{given by} \;\;(u,v)*(w,t)=(uw-vt,vw+ux)
\end{array}
\right.
$$
In our case, the complexification of $M_D$ induced by $T$ occurs only on the
subspace of $M_D$ orthogonal to the space-time $V_{4}$ which remains real and
preserves its light cone structure. The resulting complex manifold ${\cal M}
_D/ \C$, defines a ''complex embedding'' of a real space-time.

\hfill{\rule{2mm}{2mm}

As a classic example of the  signature change problem consider two well known
embeddings of the Schwarzschild space-time in six dimensional pseudo Euclidean
flat spaces:\\
Kasner~\cite
{Rosen:1}: $\;\;\;\;\;K:V_4\rightarrow {\cal M}_6\;\;\;ds^2=\;\;d{\cal Y}
_1^2+d{\cal Y}_2^2-d{\cal Y}_3^2-d{\cal Y}_4^2-d{\cal Y}_5^2-d{\cal Y}_6^2$,
\vspace{2mm}\\
Fronsdal~\cite{Fro:1}$:\;\;F:{\cal V}_4\rightarrow {\cal M}
_6^{\prime }\;\;\;-ds^2=d{{\cal Y}^{\prime }}_1^2-d{{\cal Y}^{\prime }}_2^2-d{
{\cal Y}^{\prime }}_3^2-d{{\cal Y}^{\prime }}_4^2-d{{\cal Y}^{\prime }}_5^2-d
{{\cal Y}^{\prime }}_6^2$,\\ given by (here we assume mass units such that $
2m=1$):
$$
K\left\{
\begin{array}{l}
{\cal Y}_1=(1-1/r)^{1/2}\mbox{cos}t \\ \vspace{1mm}{\cal Y}_2=(1-1/r)^{1/2}
\mbox{sin}t \\ \vspace{1mm}{\cal Y}_3=f(r),\;\;(df/dr)^2=\frac{1+4r^3}{
4R^3(r-1)} \\ \vspace{1mm}{\cal Y}_4=r\mbox{sen}\theta \mbox{sin}\phi  \\
\vspace{1mm}{\cal Y}_5=r\mbox{sin}\theta \mbox{cos}\phi  \\ \vspace{1mm}
{\cal Y}_6=r\mbox{cos}\theta \vspace{1mm}
\end{array}
\right. \;\;\;\;\mbox{and}\;\;\;F\left\{
\begin{array}{l}
{\cal Y}_1^{\prime }=2(1-1/r)^{1/2}\mbox{sinh}(t/2) \\ \vspace{1mm}{\cal Y}
_2^{\prime }=2(1-1/r)^{1/2}\mbox{cos}h(t/2) \\ \vspace{1mm}{\cal Y}
_3^{\prime }=g(r),\;\;(dg/dr)^2=\frac{(r^2+r+1)}{r^3} \\ \vspace{1mm}{\cal Y}
_4^{\prime }=rsen\theta \mbox{sin}\phi  \\ \vspace{1mm}{\cal Y}_5^{\prime }=r
\mbox{sin}\theta \mbox{cos}\phi  \\ \vspace{1mm}{\cal Y}_6^{\prime }=r
\mbox{cos}\theta \vspace{1mm}
\end{array}
\right.
$$
In the first case we have two time-like dimensions while in the second case
we have only one (we are using $-ds^2$ instead of $ds^2$. Both correspond to
the same space-time, except for a difference in topology: In $K$, the
space-time extends only to $r=1$, while in $F$ it extends to r=0. The second
embedding corresponds in fact to Kruskal's space-time, or the maximal
analytic extension of the Schwarzschild space-time.

Notice that the embedding defined by Kasner is not causal. Any curve in
the plane $(
{\cal Y}_{1},{\cal Y}_{2})$ with a parameter range greater than $2\pi$ is
closed in that embedding \cite{Fro:1}.
Since we are required to perform genuine non-local experiments to
apply the equivalence principle and to distinguish causal and non causal
propagations, we cannot rely on the implicit function theorem alone to
characterize an embedding properly.
Unless he remains strictly local, an observer in the space-time would be able
to detect if his space-time is embedded or not simply by observing a classical
breaking of causality.
In essence we are saying that the Kasner embedding cannot be used as a physical
embedding.
Nonetheless, the Schwarzschild space-time can be seen as a subset embedded
in Kruskal's space-time defined by an extension map $\Psi$ \cite{Hawking:1}.
That is, there is a third embedding of Schwarzschild's space-time,
given by composite map $Fo\Psi $. This  embedding is  consistent with the
one defined by Fronsdal and it has the appropriate signature.
\begin{figure}[tbh]
\begin{picture}(50,50)(0,0)
\thinlines
\put(45,16){\vector(1,0){33}} 
\put(45,33){\vector(1,0){33}} 
\put(45,33){\vector(0,-1){17}}
\put(78,33){\vector(0,-1){17}}
\put(45,33){\vector(2,-1){33}}
\it
\put(40,13){$V'_{4}$}
\put(38,33){$V_{4}$}
\put(78,33){${\M}_{D}$}
\put(78,14){$\M'_{D}$}
\put(60,9){$F$}
\put(60,35){$K$}
\put(38,23){$\Psi$}
\put(58,25){$Fo\Psi$}
\put(82,23){$T$}
\label{fig:K}
\put(20,0){\small Fig.2: Two embeddings of Schwarzschild space-time}
\end{picture}
\end{figure}

The matrix representing $T$ is
$$
(T^{A}{}_{B})= \left (
\begin{array}{cc}
a & b \\
c & d
\end{array}
\right ) .
$$
Replacing in (\ref{eq:T}) with  $g_{55}=1,\;g_{66}=-1$ and $g'_{55}=1,\;
g'_{66}=1$,  we obtain
$$
a^{2} +b^{2} =1,\; c^{2}+d^{2}=-1,\; ac+bd=0,\; (ad-bc)^{2}=-1
$$
One possible solution is $a=d=0,\; c=i,\; b=1 $, so that $T$ is indeed complex.

The change of signature of $\M_{D}$ may have some topological consequences.
This can be seen by taking the embedding diagrams for Schwarzschild (Kasner)
and Kruskal (Fronsdal)
space-times:
\begin{figure}[tbh]
\begin{picture}(50,70)(0,0)
\thinlines
\put(30,30){\circle{15}}
\put(30,45){\line(0,-1){30}}
\put(15,30){\line(1,0){30}}
\put(100,30){\circle{15}}
\put(100,45){\line(0,-1){30}}
\put(85,30){\line(1,0){30}}
\put(100,30){\line(1,-1){15}}
\put(100,30){\line(-1,1){15}}

\put(100,30){\line(-1,-1){15}}
\put(100,30){\line(1,1){15}}
\put(57,30){\vector(1,0){13}}
\it
\put(30,15){$V_{4}$}
\put(100,15){$V_{4}'$}
\put(117,30){$r$}
\put(45,30){$r$}
\put(28,12){$2m$}
\put(118,45){$2m$}
\put(15,43){$\R^{2}_{II}$}
\put(40,43){$\R^{2}_{I}$}
\put(63,34){$\Psi$}
\put(40,12){$\M_{D}\; (++----)$}
\put(117,12){$\M'_{D}\;(+-----)$}
\put(40,0){\small Fig.3: Topology change with signature change}
\end{picture}
\end{figure}
In figure 3, the circle in the left represent an open sphere $S^2$ which
intersects the semi planes $\R_I^2$ and $\R_{II}^2$ ,
excluding the plane $r=2m=1$. The corresponding topology is then $
(\R_I^2\cup \R_{II}^2)\times S^2$. On the
other hand, in the right hand side the topology is $\R^2\times S^2$
\cite{Brans:1}.

The above result shows that by use of complex embeddings it is possible to
preserve the space-time signature while altering only the signature of the
embedding space. For example, we may use complex transformations to make ${\cal
M}_D$ truly Minkowskian (with just one time like dimension) and use it as a
fixed background in a canonical quantization procedure while keeping intact the
space-time signature. In this case the  group  of  rotations of the normal
vectors is $SO(D-4)$ whose importance  will be seen  in the next secion.
This  may be relevant for the recent debate on the need or
not of changing the space-time signature to apply path integrals in quantum
cosmology (see e.g.  \cite{Hor:1},\cite{Ellis:1},\cite{Cecile:1}), provided it
could be made dynamical. That is, considering the embedding  equations as
part of the dynamical equations, together with Einstein's equations. In this
case, any 4-surface of  decontinuity of the  second fundamental form $b_{ij}$
may induce  a classical change of signature of  $\M_{D}$ in  a process
analogous
to that described in \cite{Ellis:1} and  eventual topological changes
\cite{Emb:1}. To see how this dynamics  takes place we need an
easier way to interpret the fundamental equations \rf{GCR} which control the
extrinsic curvature  of the space-time.

\section{The Twisting Connection}
In the following  we consider  the  embeddings  of a given space-time in a real
space with the same signature\footnote{We consider as  equivalent
signatures which differ only by a factor $-1$, or  by   a mere  relabelling
of the embedding coordinates} $p+q$. Therefore the  signature preserving
symmetery  is $T=SO(p-3,q-1)$.

The fundamental theorem of sub manifolds says that given the symmetric
tensor $g_{ij}$, $D-4$ tensors $b_{ijA}$ and $(D-4)(D-5)/2$ vectors $A_{iAB}$
satisfying (\ref{eq:GCR}) then there is a 4-dimensional sub manifold of a
flat space ${\cal M}_D$ which has $g_{ij}$ as its metric, $b_{ijA}$ as its
second fundamental form and $A_{iAB}$ as its twisting vector.

When we apply this theorem to a space-time, we have an embedded manifold
which acts as the arena for low and high energy phenomena. We could well ask
what  prevents the space-time from ``diluting'' into the ambient
space. That is, why it holds together as a four-dimensional sub manifold of
$\M_{D}$ ?
Put in another way,  a high energy particle collision or  a pair creation
could in principle eject a particle from space-time into the ambient space.
However at the current energy level this is not observed. It appears that
at this level of energy the  space-time submanifold is stable.
In the following we show that the twisting vector  $A_{iAB}$ may play a role in
that dynamics.

Since we are not assuming any external forces acting on the space-time, we may
take those curves as geodesics of ${\cal M}_D$. For each direction $N_A$ we
may define the parameter $x^A$ so that the geodesic coordinates are ${\cal Z}
^\mu ={\cal Z}^\mu (x^i,x^A)$:
\begin{equation}
\frac{\partial ^2{\cal Z}^\gamma }{\partial {x^{A}}^{2}}+\Gamma _{\alpha \beta
}^\gamma \frac{\partial {\cal Z}^\alpha }{\partial x^A}\frac{\partial {\cal Z
}^\beta }{\partial x^A}=0.
\end{equation}
For simplicity but without loss of generality, let us take ${\cal Z}^\mu $
as geodesic coordinate of ${\cal M}_D$ (actually, since ${\cal M}_D$ is flat
we may take Cartesian coordinates). In this case $\Gamma _{\alpha \beta
}^\gamma =0$ and we  may write ${\cal Z}^\mu (x^i,x^A)=\Y^\mu (x^i)+x^AN_A^\mu
$. The metric  of ${\cal M}_D$ in the Gaussian coordinate system
$(x^{i},x^{A})$ is
\begin{equation}
\label{eq:ME}{\cal G}_{\alpha\beta}^{\prime }=
{\cal Z}^\mu {}_{\alpha}{\cal Z}^\nu {}_{\beta }{\cal G}_{\mu \nu }=\left(
\begin{array}{cc}
\tilde g_{ij}+x^Ax^Bg^{MN}A_{iMA}A_{jNB} & g_{iA} \\
g_{jB} & g_{AB}
\end{array}
\right)
\end{equation}
where we have denoted
$$
\tilde g_{ij}=g_{ij}-2x^Ab_{ijA}+x^Ax^Bg^{mn}b_{imA}b_{jnB},\;\;\mbox{and}
\;\;g_{iA}=x^MA_{iMA}.
$$
Now let us consider a remarkable property of the twisting vector:

\begin{theorem}
Under an infinitesimal pseudo rotation of the normal vectors $N$, the
twisting vector transforms as:
\begin{equation}
\label{eq:gauge}A_{iAB}^{\prime }=A_{iAB}-f_{AB\;MN}^{EF}A_{iEF}\Theta
^{MN}-\Theta _{AB\,,i}
\end{equation}
where $f_{AB\;MN}^{EF}$ denote the structure constants and $\Theta^{MN}$ denote
the parameters of the group $SO(p-3,q-1)$.
\end{theorem}
{}From (\ref{eq:ME}) we may express the torsion vector as $A_{iAB}={\partial
{\cal G}_{iA}}/{\partial x^{A}}$. Therefore under an
infinitesimal transformation of $SO(p-3,q-1)$:
$$
x'^{i} =x^{i},\;\;\; x'^{A}=x^{A}+\xi^{A}
$$
Keeping only the linear terms in $\xi$, the infinitesimal transformation of
$A_{iAB}$ is:
$$
A^{\prime}_{iAB}=\frac{\partial {\cal G}^{\prime}_{iA}}{\partial
x^{B^{\prime}}}=
(\delta^{M}_{A}-\xi^{M}_{A})\frac{\partial}{\partial x'^{M}}\left (
(\delta^{\mu}_{i}
-\xi^{\mu}_{,i}) (\delta^{\nu}_{A}-\xi^{\nu}_{,A}) {\cal G}_{\mu\nu}\right) .
$$
Since  $\xi^{k}=0$ and $\xi^{A}=\Theta^{A}_{M}(x^{i})x^{M}$ we end up with
\begin{equation}
A_{iAB}^{\prime }=A_{iAB}-2g^{MN}A_{iM[A}\Theta _{NB]}-\Theta^{M}{}_{A,i}
g_{MB} .
\end{equation}
The Lie algebra of $SO(p-3,q-1)$ with generators $L^{AB}$ is given by
$[L^{MN},L^{PQ}]=f_{AB}^{MNPQ}L^{AB}$, where
\begin{equation}
f_{AB}^{MNPQ}=4\alpha \delta _A^{[N}g^{M][P}\delta _B^{Q]}\;\; \mbox{and}\;\;
f_{AB MN}^{PQ}=4\alpha \delta_{[A}^{P}g_{B][M}\delta_{N}^{Q}
\end{equation}
where $\alpha $ is a some normalization constant. Therefore,
$$
A_{iAB}^{\prime }=A_{iAB}-2A_{iEF}\frac 1{4\alpha }f_{AB\,MN}^{EF}\Theta
^{MN}-\Theta^{M}{}_{A} g_{MB}.
$$
Hence, for $\alpha =1/2$ we obtain (\ref{eq:gauge}), which is the
same as the transformation of  a gauge potential in Yang-Mills
theory, with gauge group  $SO(p-3,q-1)$.

\hfill{\rule{2mm}{2mm}

Defining  Lie-algebra valued "twisting" vector field
$$
A_{i}= A_{iAB}L^{AB} .
$$
The transformation (\ref{eq:gauge}) suggests that $A_{i}$ induce a
gauge-like connection in  $V_{4}$, the twisting connection, with the
corresponding ``gauge'' covariant derivative operator by
$$
D_{i} = \nabla_{i} + \beta A_{i} .
$$
where $\beta$ is another constant to be appropriately chosen, not necessarily
meaning a  coupling constant.  This  covariant derivative acts on Lie  algebra
valued functions ${\bf f}$ as $D_{i}\mbox{\bf f} = \nabla_{i}\mbox{\bf f} +
\beta [A_{i},\mbox{\bf f}]$. In particular, for scalar functions  $f$, $D_{i} f
= \nabla_{i} f$, so that $ D_{i}g_{jk}=0$.
Using tha  fact that $\nabla_{i}L^{AB}=0$, we obtain the commutator
\begin{equation}
\label{eq:D1} [D_{i},D_{j}] = \beta (\nabla_{i}A_{j}-\nabla_{j}A_{i}
+\beta [A_{i},A_{j}]),
\end{equation}

Next we consider the Clifford algebra associated with the metric $g^{AB}$
defined by
$$
E^AE^B+E^BE^A=2g^{AB}E^0
$$
where $E^0$ is the identity element $E^AE^0=E^0E^A=E^A$. This algebra is
closely related with the isometry group of $g^{AB}$. In fact, if the Lie
algebra of this group is generated by $L^{AB}$, then
\cite{Chevalley:1}
\begin{equation}
L^{AB}=\frac 1\gamma [E^A,E^B],
\end{equation}
where again $\gamma $ is another scale  constant to be chosen. The indices A,
B,...
are raised and lowered with $g_{AB}$ and $g^{AB}$ such that $E^A=g^{AB}E_B$.
Therefore given the coefficients of the second fundamental form $b_{ijA}$, we
may define the Clifford algebra valued tensors $b_{ij}=b_{ijA}E^A $.

\begin{theorem}
If  $F_{ij}$ is the  curvature  associated with the twisting connection, then
Codazzi's and Ricci's equations  are respectivelly  equivalent to
\be
D_{[k}b_{ij]}=0 \label{eq:Codazzi}
\te
\be
F_{ij}  =  -2g^{mn}b_{m[i}b_{j]n}   \label{eq:Ricci}
\te

\end{theorem}

In fact, since $D_i$ and $[D_i,D_j]$ are Lie-algebra valued functions, we may
write  $[D_i,D_j]=[D_i,D_j]_{AB}L^{AB},$ where we have
denoted (from \rf{D1})
\begin{equation}
\label{eq:commut}[D_i,D_j]_{AB}=\beta \left( \nabla _iA_{jAB}-\nabla
_jA_{iAB}+\beta A_{iMN}A_{jPQ}f_{AB}^{MNPQ}\right) .
\end{equation}
{}From the definition of structure constants it follows that
$$
f_{AB}^{MNPQ}L^{AB}=[L^{MN},L^{PQ}]_{AB}L^{AB}
$$
Therefore, (\ref{eq:commut}) may be written as
\begin{equation}
\label{eq:commutator}[D_i,D_j]_{AB}=2\beta (\nabla _{[i}A_{j]AB}+\beta
g^{MN}A_{[iMA}A_{j]NB})
\end{equation}
Comparing the right hand side of this expression to the left hand side of
Ricci's equation in (\ref{eq:GCR}) we obtain with $\beta =-1$,
$D_{i}=\nabla_{i}-A_{i}$  and
\be
[D_{i},D_{j}]= g^{mn}b_{m[jA}b_{i]nB}\frac{1}{\gamma}[E^{A},E^{B}]=
\frac{4}{\gamma}g^{mn}b_{[jm}b_{i]n} \label{eq:DIDJ}
\te
To complete the demonstartion, introduce the notation
\[
D_{kA}^{N} =\delta_{A}^{N}\nabla_{k} b_{jiN}-g^{MN}A_{kAM}.
\]
Then the second equation \rf{GCR}  can be written as
\be
D_{[kA}^{N}b_{j]iN}=0, \label{eq:DkAN}
\te
On the other hand, using the definition of $D_{i}$, the gauge covariant
derivative of $b_{ij}$ is given by
\begin{equation}
\begin{array}{l}
D_kb_{ij}=\nabla_{k} b_{ij}-[A_k,b_{ij}] \label{eq:Db}
\end{array}
\end{equation}
but, we can easily see that
$$
[A_k,b_{ij}]=A_kb_{ij}-b_{ij}A_k= \frac{8}{\gamma}g^{AB}A_{kCB}b_{ijA}E^C
$$
Consequently,
\[
D_{k}b_{ji} =\left(  \delta_{A}^{M} \nabla_{k}
-\frac{8}{\gamma}g^{MN}A_{kAN}\right) b_{ijM}E^{A}
\]
Comparing with \rf{DkAN} it follows that for  $\gamma =8$  we obtain
Codazzi's equation \rf{Codazzi}:
$$
D_{[k}b_{ij]}=D_{[kA}^{M}b_{ij]}ME^{A}=0
$$
Finally, the curvature  associated with  $A_{i}$ is  $F_{ij}=[D_{i},D_{j}]$,
so that   from \rf{DIDJ}  we obtain \rf{Ricci}

\hfill{\rule{2mm}{2mm}

As we see, Gauss and Ricci's equations are equivalent in the sense that the
curvature tensors of the Levi-Civita and  twisting connections in terms of the
variable $b_{ij}$, which acts as a source field subjected to Codazzi's
equation.

For completeness we may also write Gauss equation in the same  algebraic form.
This is easily accomplished using the definition of $E^A$ in the first equation
of \rf{GCR}, obtaining
$$
R_{ijkl}E^{0} =  b_{i[k}b_{l]j} -b_{j[k}b_{l]i}
$$
We conclude that the conditions for the embedding of a space-time
may be compatible with with the physics of the space-time physics, provided
the  integrability  conditions  are included as  part of the dynamics
and with the addoption of the principle of economy of dimensions.

The hidden internal indices $A,B,..$ in the algebraic  form of the
equations \rf{Codazzi},\rf{Ricci} and \rf{Gauss}, merely reflect the
degrees of freedom for the embedding which is defined up to a transformation
of the normal vectors. As such they do not affect the number of independent
equations.
To understand the space-time as a four-dimensional submanifold and why
it stays like that, depends on further understanding of $b_{ij}$ and $A_{i}$
as  physical fields in addition to the metric (the gravitational field).
This will be dealt with in a subsequent paper.

\end{document}